\documentclass{eptcs}
\providecommand{\event}{ADG 2021} 
\usepackage{url}
\usepackage{graphicx}
\usepackage{framed}
\usepackage{breqn}
\usepackage{amsfonts}
\usepackage{listings}
\usepackage{color}

\def\orcidID#1{\unskip$^{\mbox{\href{https://orcid.org/#1}{\scriptsize{[#1]}} }}$}

\title{GeoGebra Discovery in Context}
\author{Zolt\'an Kov\'acs\orcidID{0000-0003-2512-5793}
\institute{The Private University College of Education of the Diocese of Linz \\
Linz, Austria}
\email{zoltan@geogebra.org}
\and 
Tom\'as Recio\orcidID{0000-0002-1011-295X}
\institute{Universidad Antonio de Nebrija \\ Madrid, Spain}
\email{trecio@nebrija.es}
\and 
M.~Pilar V\'elez\orcidID{0000-0002-5724-4300}
\institute{Universidad Antonio de Nebrija \\ Madrid, Spain}
\email{pvelez@nebrija.es}
}

\def\titlerunning{GeoGebra Discovery in Context}
\def\authorrunning{Kov\'acs, Recio \& V\'elez}

\begin{document}

\maketitle              

\begin{abstract}
In our contribution we will reflect, through a collection of selected examples, on the potential impact of the \emph{GeoGebra Discovery} application on different social and educational contexts.


\end{abstract}

At ADG 2018 and AISC 2018 \cite{adg-ag-paper,aisc2018}  we reported about the initial steps of our work, dealing with extending the capabilities of the already performing GeoGebra automated reasoning tools towards achieving a sort of \emph{mechanical
geometer} program, that would not require human intervention for the finding of a large collection of geometric properties on a given figure. Currently, we are happy to announce that this has been accomplished (in a reasonable level of success) through  the \emph{GeoGebra Discovery} program.
 
Our contribution to this ADG21 does not deal with the technical characteristics of  \emph{GeoGebra Discovery} (that are the subject to a different submission), but with the reflection---through a collection of relevant examples---of the potential applications of this program in the educational context, among others.

Let us recall that currently  \emph{GeoGebra Discovery}  offers the user the possibility of accomplishing a rich variety of geometric tasks,
some of them well-known to the ADG community, some of them, we think, maybe quite new:
\begin{itemize}
\item automatically proving the truth or failure of a given statement (\textbf{Prove} and \textbf{ProveDetails} commands, \textit{improved}),
\item automatically discovering how to modify a given figure so that a wrong
statement becomes true (also called \textit{automated discovery}, \cite{RecioVelez1999}, \textbf{LocusEquation} command, \textit{improved}),
\item automatically discovering the relation holding
among some concrete elements of the given figure (\textbf{Relation} command, \textit{improved}, \textit{new features}),
\item automatically discovering ``all
statements holding true involving one element in the figure selected by the user'' (\textbf{Discover} command, \textit{new}),
\item automatically discovering ``all'' statements 
of a certain kind (perpendicularity, parallelism, $\ldots$) involving all the elements
of a given figure (available as a web application, see \cite{adg-ag-paper,aisc2018}, \textit{improved}),
\item improved features for locus computation (automatically adding some non-degeneracy conditions, see \cite{cicm2019}),
and improvements for envelope computation (\textbf{Envelope} command).
\end{itemize}

Moreover, we have improved some of these commands by including the possibility to deal
with geometric properties holding over the real numbers (see \cite{Wolfram2019}, \cite{scsc-2020}). 

\begin{figure}[h!]
\begin{center}
\includegraphics[width=0.6\textwidth]{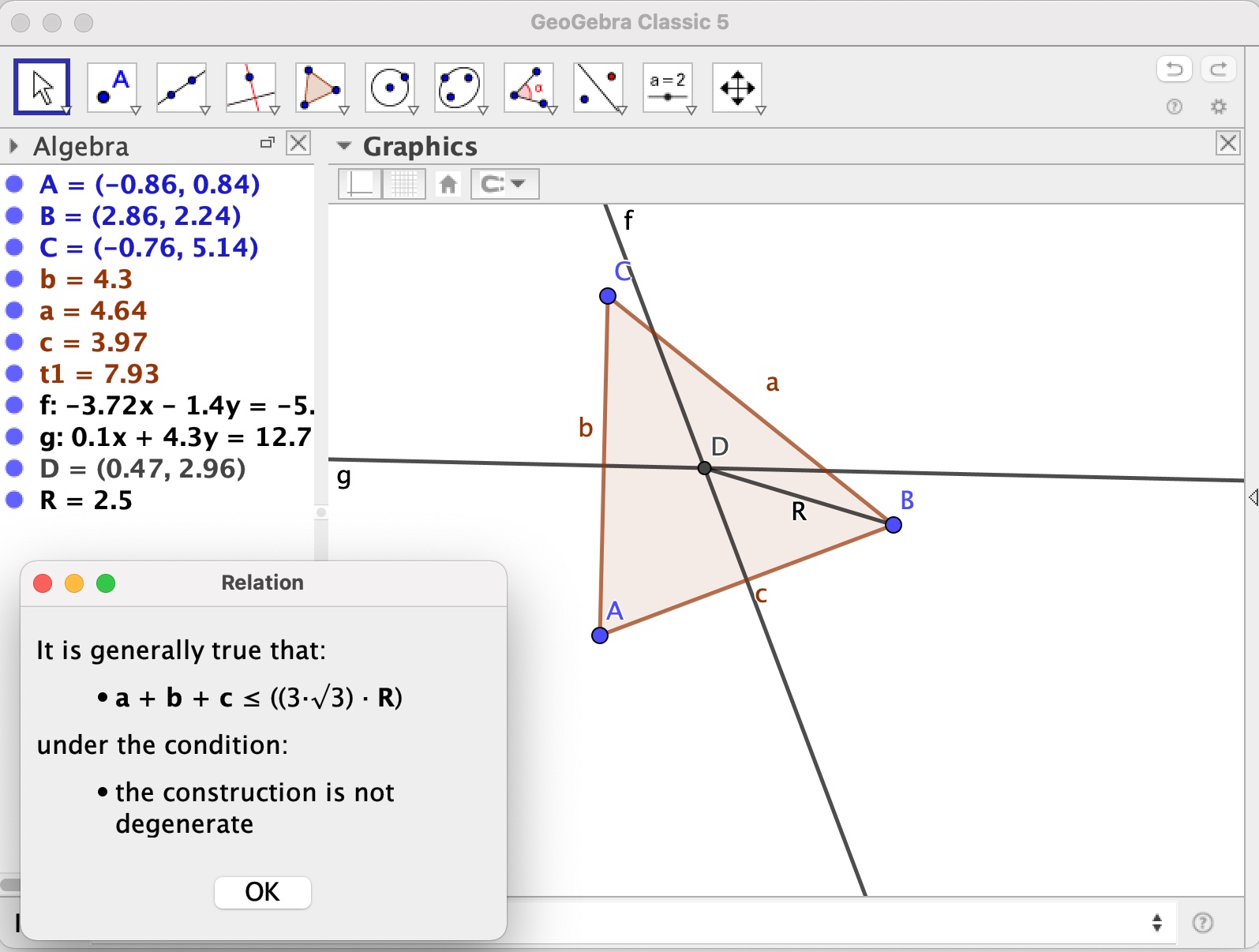}
\caption{
On an arbitrary triangle, the sum of the lenghs of sides $a+b+c$ is compared with the radius $R$ of the circumcircle.}
\label{fig1}
\end{center}
\end{figure}

Let us remark that some of these tasks, are, as
far as we know, quite unique in the context of automated reasoning, and
they were just sketched in our presentations at \cite{adg-ag-paper} and \cite{aisc2018}.
For the performance of some of these
tools we refer to \cite{ijtme2018,BKR2020,KR2020,ELRV,KRV2021}.

\begin{figure}[h!]
\begin{center}
\includegraphics[width=0.7\textwidth]{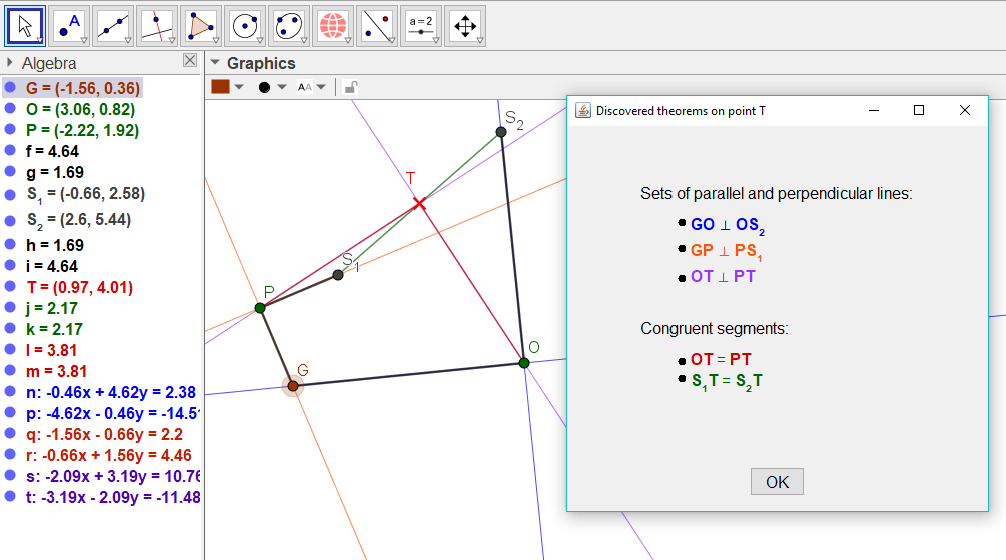}
\caption{
Two GeoGebra windows after selecting the tool \textit{Discover} and clicking at point $T$ (equivalently, introducing \textit{Discover(T)}
in the command line). The pop-up window lists different geometric theorems involving $T$ in different colors,
and the Graphics Window displays, with the same colors, the geometric objects involved in these relationships.}
\label{fig2}
\end{center}
\end{figure}

As it is well known, the mathematics  behind the \emph{mechanical geometer}
involves the translation of the geometric facts into a collection of
polynomial equations and inequations, and the corresponding manipulation
by means of (complex or real) computational algebraic geometry algorithms
developed by the authors, involving Hilbert dimension,  ideal elimination
and saturation computations  using the free computer algebra software
Giac \cite{giac}, embedded in GeoGebra
for Gr\"obner Bases computations. Moreover, some of the new features require 
the use of real algebraic geometry algorithms, like the
Cylindrical Algebraic Decomposition, freely available through Wolfram's \textit{Mathematica}
for the \textit{Raspberry Pi} machines, see \cite{Wolfram2019}.

In our talk we will present a collection of several examples
reflecting on the potential impact of the above mentioned improvements and new features of
the current version of \emph{GeoGebra Discovery}. 

Some of the examples show how \emph{GeoGebra Discovery} is able to find out---without requiring the user to guess the complicated involved formula---some inequality among different  elements of a triangle
(e.g.~Inequality 5.3, from the classical book \cite{Bottema69}), see Figure \ref{fig1}. Other examples will deal
\begin{itemize}
\item with the automatic solution of the well known pirate's ``Treasure Island Problem''  
\cite{Wilson1997}
 asking \emph{GeoGebra Discovery}  to find out whatever properties of the initial  treasure locus, to ``inspire'' the researcher,
\cite{MEIA_MEDE_KRV} (see Figure \ref{fig2}),

\item or with the generalization of a paradigmatic problem posed at the ICMI study ``School Mathematics in the 90's''
\cite{HowsonWilson1986} concerning the partition of the diagonal $AD$ of a square $ABCD$ in different segments, described by the intersection of the diagonal with lines from $D$ to the points $A_i$ in side $AB$, after dividing this side in $n$ equal parts (here the role of \emph{GeoGebra Discovery}  is rather behaving like a kind of oracle, helping the user to establish different conjectures, \cite{MEIA_MEDE_KRV} (see Figure \ref{fig3}),

\begin{figure}[htb]
\begin{center}
\includegraphics[width=0.4\textwidth]{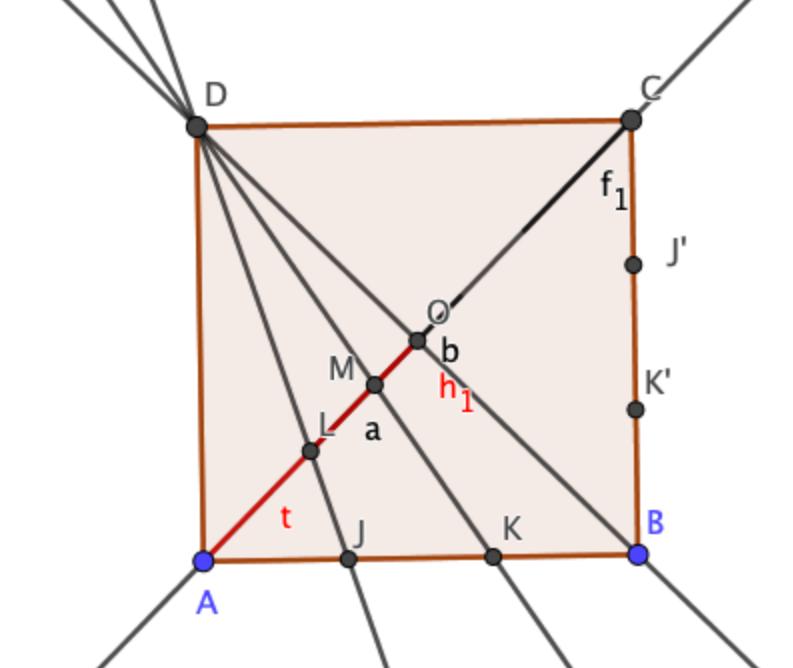}
\caption{
Kuwait ICMI Study question b) for $n = 3$. By symmetry, only half of the construction is shown. }
\label{fig3}
\end{center}
\end{figure}

\item or competing with Grade 11 students in the context of learning trajectories concerning proof,  as in Clough's conjecture, described by de Villiers \cite{DeVilliers2012} as an  ``illustration of the explanatory and discovery functions of proof with an original geometric conjecture made by a Grade 11 student'', where the role of the improved \textit{Relation, Compare} commands is crucial (see Figure \ref{fig4}),

\begin{figure}[htb]
\begin{center}
\includegraphics[width=0.8\textwidth]{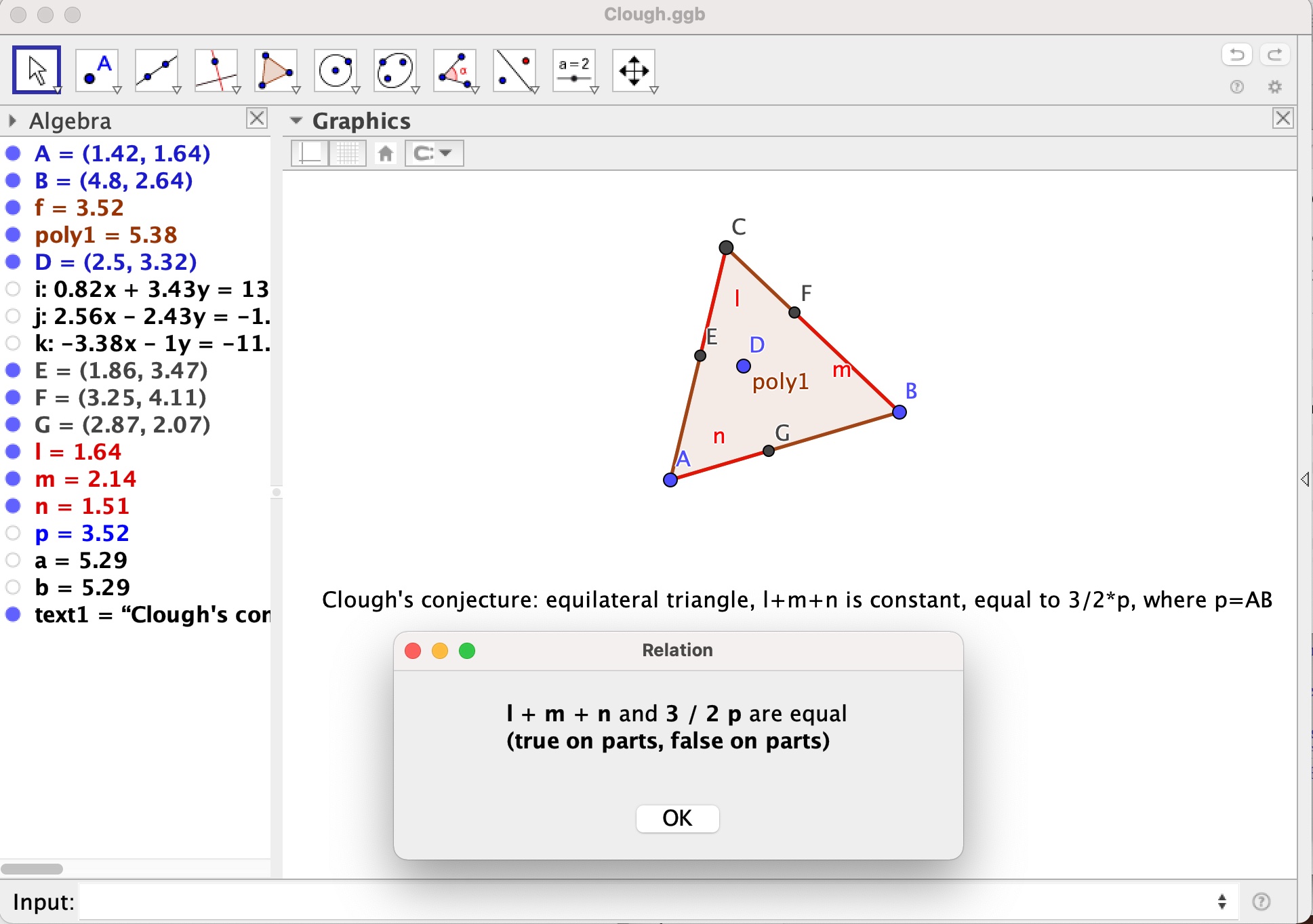}
\caption{
On an equilateral triangle, given an arbitrary point $D$ and the feet $E, F, G$ of the perpendiculars from $D$ to the sides, sum of segments $EC+FB+GA $ is equal to 3/2 of  side $AB$. }
\label{fig4}
\end{center}
\end{figure}

\item ditto, with Euclid!, trying to formulate precisely how to describe more general triangles  (besides right triangles) where the Altitude Theorem (\url{https://en.wikipedia.org/wiki/Geometric_mean_theorem}) holds (here the role of \emph{GeoGebra Discovery} is rather that of providing a correct hypothesis for a given thesis, \cite{ELRV}, see Figure \ref{fig5}),

\begin{figure}[htb]
\begin{center}
\includegraphics[width=0.6\textwidth]{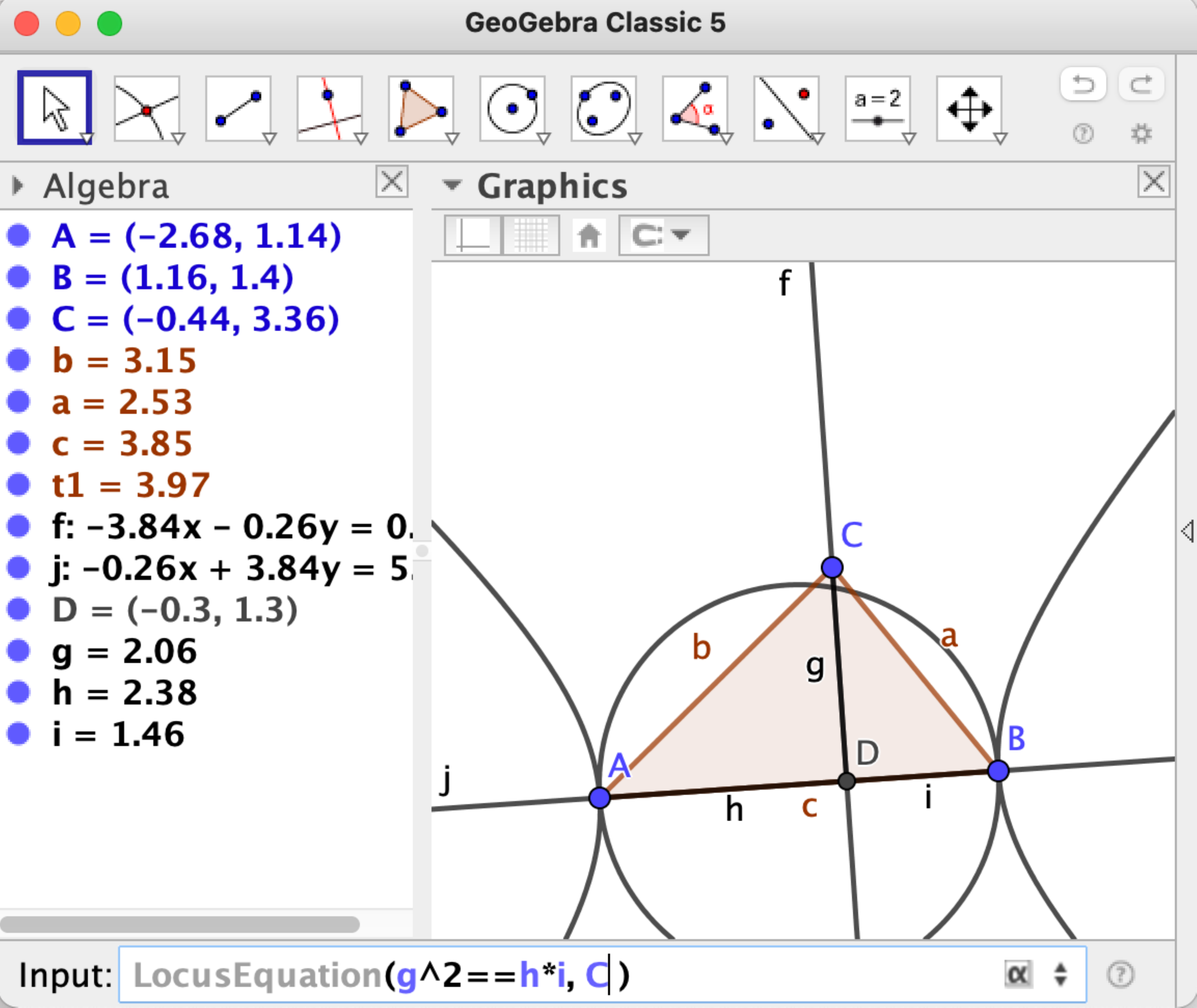}
\caption{
Extended locus for vertex $C$ so that the geometric mean theorem holds.}
\label{fig5}
\end{center}
\end{figure}

\item solving, by means of the \textit{Discover} command, some typical problem posed by university geometers in journal of math education and technology \cite{KR2021} (see Figure \ref{fig6}).

\begin{figure}[ht!]
\begin{center}
\includegraphics[width=0.7\textwidth]{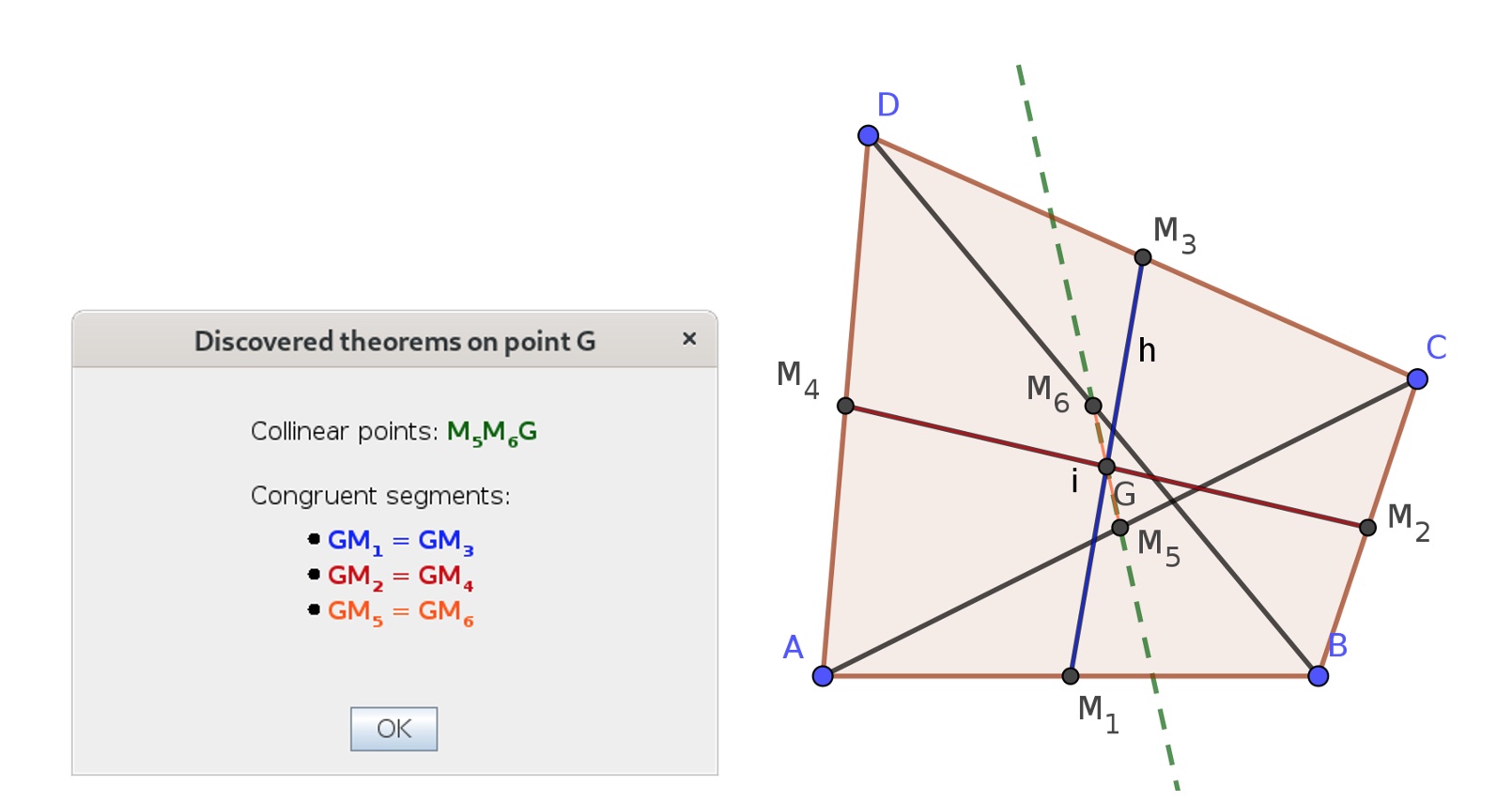}
\caption{
GeoGebra Discovery finds the solution to this problem: \textit{Let $M_1, M_2, M_3, M_4, M_5, M_6$ be the midpoints of the edges $AB, BC, CD, DA, AC, BD$. Prove that the segments $M_1M_3, M_2M_4, M_5M_6$ are concurrent in a point $G$ that bisects them all}.}
\label{fig6}
\end{center}
\end{figure}

\end{itemize} 

Through these examples, we will reflect on the potential impact of this \emph{GeoGebra Discovery} in the educational world, research,  as well as the posible contribution of this device to develop some \emph{automatically augmented reality} app \cite{BKRoseta}, starting with the automatic transformation---e.g.~by means of the Hough transform---of the image of a real object into a geometric figure in GeoGebra and then allowing the automatic analysis of its geometric characteristics.  
 
We think that this reflection (on the social role of devices performing ADG tasks) is particularly relevant at this moment, when the state of the art in this subject has achieved a high level of performance. We think it is important to consider questions like this one:  what is the purpose of developing more and more performing ADG programs? In what context are we interested in having software that finds, e.g.~the inequality between the sum of the sides of a triangle and the radius of the circumcircle?  We are aware of the fact that, perhaps, ``finding'' results is important and quite well achieved now by some programs, but helping  ``understanding'' them is still far from being satisfactorily addressed by the machines.   Here we want to emphasize that ``understanding'', i.e. producing ``readable'' (through steps or visual diagrams) proofs is a quite context-dependent issue: who are we thinking of, when we talk about understanding or reading? For example, the above mentioned  inequality is well beyond the school system level. 
 
In other words:  we think that there is a clear need to reflect (and to develop) applications, situations in concrete contexts were the current features are (or could be) useful---this is perhaps more urgent than to develop more and more efficient tools (this is also relevant, but we should have to know in advance  what would they be  good for$\ldots$). Reflecting on this should go in parallel with developing more performing programs.  This could be the final message of our contribution.

\section*{Acknowledgements}
The authors were partially supported by a grant PID2020-113192GB-I00 (Mathematical Visualization: Foundations, Algorithms and Applications) from the Spanish MICINN. Third author was supported as well by a grant from ''Grupos UCM 2021-910444. Geometría Algebraica y Analítica Real``.

\bibliography{references}

\end{document}